\documentclass[%
 reprint,
 amsmath,amssymb,
 aps,
]{revtex4-2}

\usepackage{graphicx}
\usepackage{dcolumn}
\usepackage{bm}

\newcommand{\R}{{\mathbb R}}

\newcommand{\be}{\begin{equation}}
\newcommand{\ee}{\end{equation}}
\newcommand{\beq}{\begin{eqnarray}}
\newcommand{\eeq}{\end{eqnarray}}

\newcommand{\la}{\label}

\begin{document}


\title{Hartle-Hawking's vacuum is full of Vilenkin's universe-antiuniverse pairs}

\author{Salvador J. Robles-P\'erez}
\email{sarobles@math.uc3m.es}
\affiliation{Departamento de Matem\'aticas, Universidad Carlos III de Madrid. Avda. de la
	Universidad 30,  28911 Legan\'es, Spain.}

\date{\today}

\begin{abstract}
Within the third quantisation formalism, we find solutions of the Wheeler-DeWitt  in terms of two sets of modes that can be identified with the Hartle-Hawking's no boundary condition and with the Vilenkin's tunneling boundary condition, respectively. The two sets of modes are related by a Bogolyubov transformation so the no boundary vacuum state turns out to be equivalent to a thermal distribution of tunnelling positive and negative modes representing, in a relative sense, universes and antiuniverses. 
\end{abstract}

\maketitle


\section{Creation of universes in pairs}\label{sec02}

\begin{figure}
\centering
\includegraphics[width=7 cm]{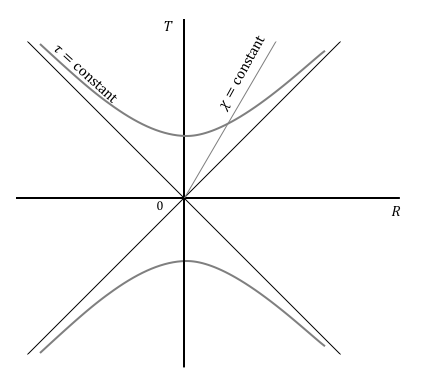}
\caption{The space of three metrics, $M$,  turns out to be a particular coordination of upper (lower) light cone of a $6$-dimensional Minkowski space. Every point in the $T,R$ plane is a four-sphere of unit radius. Lines of constant $\tau$ are lines of constant volume of the spatial sections of the spacetime (with different shapes). Lines of constant $\chi$ correspond to different volumes of the same shape (a scaling universe).}
\label{figure01}
\end{figure}

It is  known for a long time \cite{DeWitt1967} that the Wheeler-DeWitt equation (WDWE) has the structure of a Klein-Gordon equation, with the wave function $\phi(h_{ij})$ playing the role of the scalar field and the space of $3$-dimensional Riemann metrics, $M \equiv {\rm Riem}(\Sigma)$, playing the role of the subjacent space where the field \emph{propagates}. A symmetric $3$-dimensional metric has only $6$ independent components so, $M \sim \R^6$. Furthermore, with the so-called DeWitt metric that arises in the context of general relativity, it turns out that the space $M$ is a $5+1$ dimensional space with a $1$  \emph{time-like} dimension and a $5$ dimensional \emph{space-like} orthogonal subspace, $\bar M$, with a line element in $M$ given by \cite{DeWitt1967}
\be\label{SME01}
d s^2 = - d\tau^2 + h_0^2 \tau^2 \bar G_{AB} d\bar q^A d\bar q^B ,
\ee
where, $\bar q^A$, $A=1,\ldots, 5$, are the coordinates in $\bar M$, and 
\be\label{Gbar01}
\bar G_{AB} = {\rm tr}\left( h^{-1} h_{,A} h^{-1} h_{,B} \right) \equiv h^{ij} \frac{\partial h_{jk}}{\partial \bar q^A} h^{k l} \frac{\partial h_{l i}}{\partial \bar q^B}  .
\ee
The time-like variable $\tau$ is defined as  \cite{DeWitt1967}
\be
\tau = h_0^{-1} h^{1/4} ,
\ee
with, $h_0^2 = 3/32$, and essentially represents the volume of the spatial sections of the universe ($\propto \sqrt{h}$). It turns out that $\bar M$ is a metric space with constant negative curvature given by
\be\label{SCA101}
\bar R = \bar R_{AB} \bar G^{AB} = -\frac{15}{4} \equiv - \frac{1}{a^2} .
\ee
From \eqref{SME01} and \eqref{SCA101}, it is clear that the space $M$ has the same formal structure of a Friedmann-Robertson-Walker spacetime with the hyperbolic $5$-space $H^5$ as the spatial section. In particular, it has the same formal structure than the Milne spacetime. Therefore, one can find a set of coordinates  in $\bar M$, $\bar q^A = (\chi, \theta, \phi, \psi, \zeta)$, in terms of which the metric \eqref{SME01} can be written as \footnote{A rescale, $\chi \rightarrow a^{-1}h_0^{-1} \chi$, $\theta \rightarrow a^{-1} h_0^{-1} \theta$, $\ldots$, has been made to absorb the constants $a$ and $h_0$.} 
\be\label{SME102}
d s^2 = - d\tau^2 + \tau^2 \left( d\chi^2 + \sinh^2\chi d\Omega_4^2 \right)  ,
\ee
where, $\chi \in[0,\infty)$, and $d\Omega_4^2$ is the line element on the $4$-sphere of unit radius. The Milne spacetime is a particular coordination of part of the Minkowski spacetime. It only covers the interior of the upper (or the lower, see below) light cone of the Minkowski spacetime.  Something similar occurs in $M$. Let us introduce the variables 
\be
T = \tau \cosh  \chi \ \ , \ \ R = \tau \sinh  \chi ,
\ee 
in terms of which the line element \eqref{SME102} becomes
\be\label{SME04}
ds^2 = - dT^2 + dR^2 + R^2 d\Omega_4^2 ,
\ee
with, $0 < T < \infty$ and $-\infty < R < \infty$. The metric \eqref{SME04} is nothing more than the metric of a $6$-dimensional Minkowski space, and this coordination of the Milne space only covers the upper light cone (see Fig. \ref{figure01}). The interior of the lower light cone is covered by a time reversal change of coordinates, $\tau \rightarrow -\tau$ (let us notice that the metric \eqref{SME102} is invariant under this change).

We can now develop the quantum field theory of a field $\phi(h_{ij})$ that propagates in the $6$-dimensional space $M$. With an appropriate choice of the factor ordering, the Wheeler-DeWitt equation can be written as the wave equation \footnote{Following Ref. \cite{DeWitt1967}, we are using units in which, $\hbar = c = 16\pi G = 1$. However, in the quantum development we explicitly maintain the Planck constant, $\hbar$, to make clear the semiclassical approach that we shall consider in the next subsections.},
\be\label{WDE01}
\left( -\hbar^2 \Box_q + m^2_g(q) \right) \phi(q) = 0 ,
\ee
with,
\be\label{MAS101}
m_q^2(q) \equiv 2 V = h_0^2 \tau^2  \left( 2\Lambda - \; ^{3}R \right) ,
\ee
where $\Lambda$ is the cosmological constant, which can also associated with the constant potential of the inflationary field (see, Sec. \ref{sec03}), and $^{3}R$ is the scalar curvature of the spatial section of the spacetime. In \emph{conformal time}, $\lambda = \ln \tau$, and with the rescale, $\phi = e^{-2\lambda} \tilde \phi$, the wave equation \eqref{WDE01}   becomes
\be\label{WDE02}
\left\{ \frac{\partial^2}{\partial \lambda^2} - \Box_{\bar q} + \left( \frac{m_g^2}{\hbar^2} e^{2\lambda} - 4 \right) \right\} \tilde \phi(\lambda, \bar q) = 0 ,
\ee
where,
\be\label{BOX01}
\Box_{\bar q} =  \frac{1}{\sqrt{ \bar G}} \frac{\partial }{\partial \bar{q}^A} \left(\sqrt{\bar G} \,  \bar G^{AB} \frac{\partial}{\partial \bar{q}^B} \right) ,
\ee
with, $\bar G = {\rm det}\, \bar G_{AB}$. The 'mass' $m_g$  of the field, given in \eqref{MAS101},  is not a constant. In general the ${}^3R$ curvature depends on all the components of the spatial metric and thus the space $M$ turns out to be a dispersive medium for the wave function of the universe. However, as the universe expands the curvature of the space decreases on cosmological scales and become subdominant in \eqref{MAS101} provided that the value of $\Lambda$ is large enough, as expected during inflation. In that case, we can assume the value \footnote{The factor $\hbar^2$ has been introduced for later convenience.}
\be\label{MAS03}
m^2_g(q) \approx 2 \Lambda h_0^2  \tau^2 \equiv \hbar^2 m_0^2 e^{2\lambda},
\ee 
in the Wheeler-DeWitt equation \eqref{WDE01}, and perform the quantisation in the customary way (see, for instance, Refs. \cite{Birrell1982, Mukhanov2007}). We decompose the wave function of the universe $\phi(q)$, with $q=(\lambda, \bar q)$,  in normal modes as
\be\label{FD01}
\phi(q)=\int_0^\infty d k \sum_{\vec j}  \, [ a_\textbf{k} u_\textbf{k}(q) + a_\textbf{k}^\dag u^*_\textbf{k}(q) ] ,
\ee
where, $\textbf{k} = (k, \vec j)$,
\be
u_\textbf{k}(q) =e^{-2\lambda}  \chi_{k, J}(\lambda) \mathcal Y_{J, \vec M}(\bar q)  ,
\ee
and, $\mathcal Y_{k, {\vec j}}(\bar q) $, are the  eigenfunctions of the  Laplacian defined on the $5$-dimensional hyperboloid, which satisfy \cite{Bander1966}
\be
\Box_{\bar q}\mathcal Y_{k, {\vec j}}(\bar q) = - (k^2 +4) \mathcal Y_{k, {\vec j}}(\bar q) ,
\ee
with, $ 0 < k < \infty$, and $\vec j$ denotes the $4$ indices that distinguish the four components of the generalisation of the angular momentum on the $4$ sphere. Thus,  the wave equation \eqref{WDE02}  reduces to
\be\label{QO101}
\chi_k'' + \left( m_0^2 e^{4\lambda } + k^2 \right) \chi_k = 0 ,
\ee
where, $\chi' \equiv\frac{d \chi}{d \lambda}$. The wave equation \eqref{QO101} is readily solvable in terms of Bessel functions. With the customary normalisation condition
\be\label{ORN101}
\chi_k \partial_\lambda \chi_k^* - \chi_k^* \partial_\lambda\chi_k = i .
\ee
we easily find two set of orthonormal modes given by 
\beq\label{CHI101}
\bar \chi_k(\tau) &=&  \left( \frac{4}{\pi}\sinh(\pi k) \right)^{-\frac{1}{2}} \mathcal J_{-i k }(m_0 \tau^2)  , \\ \label{CHI102}
\chi_k(\tau) &=& \frac{1}{2}  \sqrt{\frac{\pi}{2}} e^{\frac{k \pi }{2}} \mathcal H^{(2)}_{ik}(m_0 \tau^2)  .
\eeq

We have to impose some boundary condition. For this, we have to notice that the multiverse is a closed system and no external influence is expected to modify its state. Therefore, one would expect that the field that represents the whole multiply connected spacetime manifold  would remain in a steady state. It is therefore appropriate to use an invariant representation \cite{Lewis1968, *Kanasugui1995, *RP2017d}, in which the field remains in the same state along the entire evolution. In particular, it means that if the field $\phi(h_{ij})$ is in the vacuum state of an invariant representation it will remain in the same vacuum state for any value $\tau$. 

An invariant representation can be given in terms of creation and annihilation operators, $\hat b_\textbf{k}$ and $\hat b^\dag_\textbf{k}$, defined as \cite{Kim2001}
\beq\label{INV101}
\hat b_\textbf{k} &=& \frac{i}{\sqrt{\hbar}} \left( \chi_k^* \, \hat p_\phi - (\chi_k{k}^*)' \hat \phi \right)  , \\  \label{INV102}
\hat b^\dag_\textbf{k} &=& -\frac{i}{\sqrt{\hbar}} \left( \chi_k{k} \, \hat p_\phi - ( \chi_k{k})' \hat \phi \right) ,
\eeq
where, $\hat \phi$ and $\hat p_\phi$, are the operator version of the wave function and the conjugate momentum in the Schrödinger picture, respectively, and $\chi_k $ is a solution of the wave equation \eqref{QO101} satisfying the orthonormality condition \eqref{ORN101}, which ensures the usual commutation relations,
\be\label{COM101}
[\hat b_\textbf{k}, \hat b_\textbf{k}^\dag] = 1.
\ee

The condition (\ref{ORN101}) does not fix the vacuum state. For instance, the modes (\ref{CHI101}-\ref{CHI102}) define two vacuum states,  $|\bar 0_\textbf k \bar 0_{-\textbf k}\rangle$ and $| 0_\textbf k 0_{-\textbf k}\rangle$, respectively. Following the reasoning made in \cite{Birrell1982}, the state $|\bar 0_\textbf k \bar 0_{-\textbf k}\rangle$ can be identified with the conformal vacuum and the state $| 0_\textbf k 0_{-\textbf k}\rangle$ with the vacuum state of the $6$-dimensional Minkowski space. On the other hand,  the modes \eqref{CHI101} can be identified with the Hartle-Hawking no boundary wave function \cite{Hartle1983}, $\chi_{HH}$. Let us notice that in the present case, the condition $^{3}R > 2 \Lambda$ in \eqref{MAS101} defines the Euclidean region from where the universes tunnel out  to become newborn universes. The no boundary proposal would select the states that are created from the $4$-geometries that would have as the only boundary the hypersurface $\Sigma_0$ for which, $^{3}R = 2 \Lambda$. As it is well known, the semiclassical mode for the Hartle-Hawking ground state becomes \cite{Hartle1983}
\be\label{HH101}
\chi_{HH} \approx 2 \cos \left( S(a) - \frac{\pi }{4} \right) \sim e^{i S(a)} + e^{-i S(a)} ,
\ee
where, $S(a) \propto a^3 \sim \tau^2$,  is the action of the spacetime and $a$ is the scale factor. On the other hand, using the asymptotic expansions of the Bessel functions \cite{Abramovitz1972}, one can check that the modes $\bar\chi_k$ become at large values of the volume $\tau$,
\be
\bar\chi_k(\tau) \propto \cos\left( m_0 \tau^2 - \frac{\pi}{4}+ \frac{ik\pi}{2}\right) ,
\ee
which agrees with $\chi_{HH}$ for the zero mode, which is the only mode computed in Ref. \cite{Hartle1983}.

From \eqref{HH101}, it is easy to see that the Hartle-Hawking state is a linear superposition of two branches, $e^{\pm i S}$. Typically, one is considered to represent an expanding universe and the other a contracting universe. Vilenkin's tunnelling boundary condition \cite{Vilenkin1982, *Vilenkin1986} imposes that the only branch that survives the Euclidean barrier is the branch that represents an expanding universe, $e^{-i S}$. Using again the asymptotic approximation of the Hankel functions, one can check that the modes $\chi_k$ in \eqref{CHI102} satisfy
\be
\chi_k(\tau) \propto e^{-i  m_0 \tau^2 } ,
\ee
in the limit of large\footnote{compared with the Planck volume} volumes. Thus, we can identify the modes $\chi_k$ with the Vilenkin's tunnelling wave function and the Minkowski vacuum state, $| 0_\textbf k 0_{-\textbf k}\rangle$, with the vacuum state of the tunnelling proposal. Let us notice that the modes $-\textbf{k}$ are given by the complex conjugated of \eqref{CHI102}, which in the limit of large $\tau$ becomes, $\chi_k \propto e^{+im_0 \tau^2}$. It means that the state, $| n_\textbf k n_{-\textbf k}\rangle$, represents the quantum state of $n$ pairs of expanding and contracting universes. In the next section, we shall see that an alternative interpretation is that it represents the state of $n$ expanding universe-antiuniverse pairs.

An interesting result that appears in the third quantisation formalism is that the two set of modes (\ref{CHI101}-\ref{CHI102}) are related by a Bogolyubov transformation, 
\be
\bar \chi_{k} = \alpha_k \chi_{k} + \beta_k \chi_{k}^* ,
\ee
where,
\be
\alpha_k = \left[ \frac{e^{\pi k}}{2 \sinh(\pi k)}\right]^\frac{1}{2} , \beta_k =  \left[ \frac{e^{-\pi k}}{2 \sinh(\pi k)}\right]^\frac{1}{2}  ,
\ee
with, $|\alpha_k|^2-|\beta_k|^2 = 1$. It means that the vacuum state of the Hartle-Hawking modes, $|\bar 0_\textbf k \bar 0_{-\textbf k}\rangle$ can be written as \cite{Mukhanov2007}
\be\label{VS01}
|\bar 0_\textbf k \bar 0_{-\textbf k}\rangle = \prod_\textbf{k} \frac{1}{|\alpha_k|^{1/2}} \left( \sum_{n=0}^\infty \left( \frac{\beta_k}{\alpha_k} \right)^n |n_\textbf{k} n_{-\textbf k}\rangle \right) .
\ee
In the no bar representation, the state \eqref{VS01} contains a number of universes given by
\be
N_k = |\beta_k|^2 = \frac{1}{e^{2\pi k} - 1} ,
\ee
which corresponds to a thermal distribution with generalised temperature
\be
T = \frac{1}{2\pi}.
\ee
Then, one can state that in the limit of large volumes, $\tau$, the Hartle-Hawking no-boundary version of the vacuum state is full of Vilenkin's pairs of universes. The result is quite general and it implies that the creation of the universes in  universe-antiuniverse pairs could be quite typical. It is formally similar to what happens in the quantum field theory of a matter field in an isotropic background spacetime, where the isotropy of the space makes that the particles are created in pairs with opposite values of the field modes, $\textbf{k}$ and $-\textbf{k}$. In the third quantisation formalism, the space-like subspace $\bar M$ is isotropic too, and the potential of the WDWE does not break this isotropy in the limit of large values of $\tau$. It means that irrespective of how the universes are initially created, they become isotropic pairs at volumes much larger than the equivalent Planck volume, which is already expected even in the first stage of the inflationary period. Thus, the possibility that our universe has been created within a universe-antiuniverse pair seems to be quite plausible.

This seems to be corroborated from a geometrical standpoint. Let us notice that the Milne spacetime can separately cover the interior of the upper and the lower light cones of the Minkowski spacetime. These two sections of the full light cone can be seen as the regions of the spacetime where propagate future oriented and past oriented particles, or particle-antiparticles, which besides turn out to be  entangled \cite{JOlson2011}. One would expect something similar in the case of the space $M$, which also covers the upper and lower half light cones of the $6$ dimensional Minkowski space. These two regions would describe entangled pairs of universes expanding and contracting universes. However, we will see in the next section that contracting universes can be seen as expanding universes with a matter field that is the charge conjugated of the field in the partner universe. Furthermore,  the value of each mode of the Fourier decomposition in \eqref{FD01} is proportional to the momentum conjugated to the variables, $\bar q^A = (\chi, \theta, \phi, \psi, \zeta)$, which are eventually related to the components of the normalised metric tensor, $\bar h_{ij} = h^{-1/3}h_{ij}$ (they are both two sets of coordinates in $\bar M$). It means that any change that is produced by the momentum associated to $+\bar{ \textbf{k}}$ in the shape of the universe with metric $\bar h_{ij}$ is being also produced in the shape of the partner universe with opposite sign, $-\bar{ \textbf{k}}$, so the parity of the two spatial sections is reversely related and so it is the relative parity of the fields that propagate in the two universes. That confirms that one of the universe of the symmetric pair is  filled with a content that is parity conjugated of the matter of the partner universe. In the next section we shall  see that they are also related by the charge conjugation operation, making plausible the interpretation of the pair of universes as a universe-antiuniverse pair.

\section{Reheating and the appearance of matter}\label{sec03}

In order to analyse the matter content of the universes after the inflationary period we have to analyse the reheating period, where the inflaton field eventually decays into the particles of the Standard Model (SM) (see, for instance, Ref. \cite{Mukhanov2008}). The formalism presented here is quite general and can be applied to different reheating scenarios.  However, for concreteness, we shall review it in the appealing  Higgs-inflation scenario \cite{Bezrukov2008}, in which the the Higgs boson of the Standard Model (SM) of particle physics is identified with the inflaton field that drives the expansion of the earliest stage of the universe, joining together the physics of particles  and cosmology.

We shall not enter into the details, which can be seen for instance in Refs. \cite{Bezrukov2008, Bellido2009}. Basically, the mechanism is based on the existence of a field with a potential term that at large values of the field presents a region of sufficient flatness to allow the universe to undergo the inflationary expansion \cite{Bezrukov2008}. However, for small values the potential turns out to be the usual potential involved in the Spontaneous Symmetry Breaking of the SM that gives masses to the fermions and the intermediate gauge bosons. Thus the masses of the SM are recovered in the low energy limit, as expected.

If one takes into account the matter fields, the total Hamiltonian constraint that gives rise the WDW equation would be,
\be\label{HC01}
\hat H_T \phi = \left( \hat H_{G} + \hat H_{HSM}\right) \phi = 0 ,
\ee
where $\hat H_{G}$ is the Hamiltonian operator that yields the WDW equation of the spacetime geometry alone \eqref{WDE01} with $\Lambda$ related to the constant part of the potential of the Higgs-inflaton field, $2 \Lambda = 2 V_0 \equiv H_0^2$, and  $\hat H_{HSM}$ is the Hamiltonian operator that contains the varying terms of the Higgs and the rest of fields of the SM as well as their corresponding potentials, including the interaction between the Higgs and the gauge and fermion fields of the SM. Following the standard procedure \cite{Hartle1990, Kiefer2007}, the wave function of the universe,  $\phi=\phi(h_{ij},\xi_0;  \varphi)$, where $\xi_0$ is the constant value of the Higgs field during the inflationary period and $\varphi$ represents the fields of the SM including the varying part of the Higgs, can be written as the product of two components, a wave function $\phi_0$ that depends only on the gravitational degrees of freedom and the constant initial value $\xi_0$, and a wave function that contains all the dependence on the fields,
\be\label{PHI00}
\phi^\pm(h_{ij}, \xi_0; \varphi) = \phi_0^\pm(h_{ij}, \xi_0) \psi_\pm(h_{ij}, \xi_0; \varphi) ,
\ee
where the two signs have been introduced for later convenience and, $\phi^+ = \left( \phi^-\right)^*$. The wave function $\phi_0$ is the solution of the WDWE of the geometrical degrees of freedom, computed in the preceding section in terms of the modes (\ref{CHI101}-\ref{CHI102}) and their complex conjugated. In general, it can be written in the semiclassical approach as
\be\la{PHI01}
\phi_0^\pm(h_{ij}, \xi_0) \propto e^{\pm \frac{i}{\hbar} S(h_{ij}, \xi_0)} .
\ee
If one introduces the wave function \eqref{PHI00} into the complete WDW equation and use the classical constraint one obtains, at order $\hbar^1$, the following equation
\be\la{SCH01}
\mp 2 i\hbar \vec \nabla S  \cdot \vec \nabla \psi_\pm = H_{HSM} \psi_\pm ,
\ee
where $\vec \nabla$ is the gradient in $M$ and the negative and the positive signs correspond, respectively, to $\phi^+$ and $\phi^-$ in (\ref{PHI00}). The Schrödinger equation for the matter fields is then obtained if one defines the (WKB) time parameter $t$ through the condition,
\be\label{WKBt01}
\frac{\partial }{\partial t} = \mp 2 \vec\nabla S \cdot \vec\nabla \equiv \mp 2 G^{\alpha \beta} \frac{\partial S}{\partial q^\alpha} \frac{\partial }{\partial q^\beta} ,
\ee
where, $q^{\alpha} = (\tau, \bar q^A)$, and $\bar q^A$ are the coordinates of $\bar M$ given in \eqref{SME01}. We have now two choices. Typically, it is chosen the positive sign in (\ref{WKBt01}) for the spacetime represented by the wave function $\phi_0^-$ and the negative sign for the spacetime represented by the wave function $\phi_0^+$. With these choice, the Schrödinger equation in the two branches turns out to be 
\be\la{SCH02}
i\hbar \frac{\partial \psi_\pm}{\partial t_\pm} = H_{HSM}(\varphi) \psi_\pm ,
\ee
where it can now be written, $\psi_\pm=\psi_\pm(t_\pm; \varphi)$. From \eqref{WKBt01}, one easily gets
\be
\frac{\partial \tau}{\partial t_\pm} = \pm 2 \frac{\partial S}{\partial \tau} ,
\ee
so the wave functions $\psi_\pm$ represent two universes, one expanding and one contracting (recall that the variable $\tau$ is proportional to the volume of the space), which from \eqref{SCH02} are both filled with matter. An alternative although equivalent interpretation is to choose the positive sign in \eqref{WKBt01} for both universes, i.e. $t \equiv t_+$. In that case, both wave functions represent expanding universes but then the corresponding Schrödinger equations for the internal fields are given by
\beq\label{FI101}
i\hbar \frac{\partial \psi_+}{\partial t} &=& H_{HSM}(\varphi) \psi_+ , \\  
- i\hbar \frac{\partial \psi_-}{\partial t} &=& H_{HSM}(\varphi) \psi_- ,
\eeq
respectively. The last of which can be written as,
\be\label{FI102}
i\hbar \frac{\partial \psi_+}{\partial t} = H_{HSM}(\bar \varphi) \psi_+ ,
\ee
where we have used that, $\psi_-^*(\varphi) = \psi_+(\bar\varphi)$. It is therefore the Schrödinger equation of a field that is  charge conjugated with respect to the field given in \eqref{FI101}. The wave functions  $\phi^+$ and $\phi^-$ represent then two expanding universes but from the point of view of the same time variable one is filled with matter and the other with antimatter, having these two concepts always a relative meaning.

At the end of the inflationary period the Higgs-inflaton field has slow rolled down the potential and it approaches the minimum of the potential located at $\chi_m$, for which $V'(\chi_m)=0$. The expansion rate of the spacetime slows down as well and the field starts oscillating around the minimum like a weakly damped harmonic oscillator with mass, $m^2=V''(\chi_m)$.  For instance, for a power-law evolution of the background spacetime the Higgs field can be written as \cite{Bellido2009}
\be\label{CHI01}
\chi(t) = \frac{\chi_{\rm end}}{Mt} \sin (Mt) ,
\ee
where, $\chi(t=0)  = \chi_{\rm end}$, is the value of the Higgs field at the end of the inflationary period, which coincides with the beginning of the appearance of the (p)reheating mechanisms ($t=0$).

Different channels can now be considered for the decaying of the Higgs field into the particles of the SM (see, Ref. \cite{Bellido2009, Koffman1997} for the details). It turns out that the perturbative decay of the Higgs field is only effective when the amplitude of the Higgs is below a critical value that depends on the mass of the final particles. This, together with the dependence of the decay rate of the Higgs into the particles of the SM makes that the Higgs needs to oscillate a large number of times before decaying into the massive gauge bosons and fermions and much more times to decay into the less massive fermions, so the perturbative decay becomes ineffective during the first oscillations of the Higgs. In that period, the most effective channel turns out to be the parametric resonance \cite{Bellido2009, Koffman1997}. This channel is enhanced by the effect of Bose stimulation so the production of fermions through this channel is highly restricted. These will be mainly produced later on through the perturbative channel or through the subsequent decay of the intermediate bosons into fermions.

Therefore, we shall mainly focus on the production of the intermediate gauge bosons, $W^\pm$ and $Z$. During the reheating period the fields of the SM acquires mass from the interaction with the Higgs, which can be approximated by \cite{Bellido2009}
\be\label{MAS01}
m_W^2 \simeq  \frac{ g_2^2 |\chi|}{4 \sqrt{6} \xi } \, , \, m_Z^2 \simeq \frac{m_W^2}{cos^2\theta_W} \kappa \, , 
\ee
where $g_2$ is the coupling of the intermediate gauge bosons and  $\theta_W$ is the weak mixing angle.  The quantisation of the intermediate gauge bosons $W^\pm$ and $Z$ follows as usual, by decomposing them into normal modes that satisfy the wave equation
\be\label{MEX02}
\ddot \varphi_k + 3\frac{\dot a}{a} \dot \varphi_k + \omega_k^2(t) \varphi_k = 0 ,
\ee
with, $\varphi\equiv W^\pm, Z$, and
\be\label{FRE01}
\omega_k^2 = \frac{k^2}{a^2} + m_\varphi^2(t) ,
\ee
where $m_\varphi$ is given by (\ref{MAS01})  with the value of the Higgs given in (\ref{CHI01}).  The time dependence of the frequency entails the production of particles. Rapidly, the intermediate gauge bosons start decaying into the fermions of the SM through their mutual interaction given by the Hamiltonian \cite{Bellido2009}
\be
H_I = -\frac{g_2}{\sqrt{2}} \left( W^+_\mu J^-_\mu + W^-_\mu J^+_\mu \right) - \frac{g_2}{\cos\theta_W} Z_\mu J^\mu_Z ,
\ee
where, $J^-_\mu \equiv \bar d_L \gamma^\mu u_L$ and $J_\mu^+ \equiv \bar u_L \gamma^\mu d_L$, are the charged currents that couple to the boson $W^+$ and to the boson $W^-$, respectively,  and the neutral current
\be
J^\mu_Z \equiv \kappa_1 \bar u_L \gamma^\mu u_L + \kappa_2 \bar d_L \gamma^\mu d_L  ,
\ee
with $\kappa_1$ and $\kappa_2$ the corresponding coupling constants. These interactions lead to the charged decays
\beq\label{D1}
W^+ \rightarrow  u + \bar d \,\, , \,\, W^- \rightarrow  \bar u +  d ,
\eeq
where $d$ and $u$ stands for the down- and up-type quarks, respectively, and similar decays can also be considered for the rest of quarks.  Analogously, we can consider the following decays in the lepton sector
\beq\label{D3}
W^+ \rightarrow  e^+ + \nu_e \,\, , \,\, W^- \rightarrow  e^- + \bar\nu_e ,
\eeq
all of them with their respective decay widths, $\Gamma_{W^\pm \rightarrow i}$. Let us then notice that an asymmetry in the decay of the Higgs into the intermediate gauge bosons would entail an asymmetry in the production of quarks and leptons and therefore an asymmetry in the creation of primordial matter during the (p)reheating period without the need of any other mechanism\footnote{Although other mechanisms of baryon asymmetry can simultaneously be present.}.

In the scenario presented in Sec. \ref{sec02} of an infinite number of pairs of universes, the Schr\"odinger equation of the fields of the SM in the two symmetric universes is given by (\ref{FI101}) and \eqref{FI102}, respectively,  with $ \psi_-^*(t, \varphi) = \psi_+(t, \bar\varphi)$. Let us focus on one of these two wave functions, say $\psi_+$. If we consider that the modes of the field are decoupled, then, the Schr\"odinger equation for the scalar field $\varphi_+$, which generically denotes any of the polarisations of the $W^\pm$ and $Z$ bosons, can be written as the product of the wave functions of the modes, i.e.
\be\label{PSI01}
\psi_+(t, \varphi) = \prod_k \psi^{(k)}_{+}(t, \varphi_k) ,
\ee
where $\psi^{(k)}_{+}(t_+, \varphi_k) $ is the solution of the Schr\"odinger equation \eqref{FI101} for each mode, whose general solution can be expressed in the basis of number eigenfunctions of the time dependent harmonic oscillator (see, for instance, Ref. \cite{Brizuela2019}). The wave function in the time reversely symmetric universe, $\psi_-(t, \bar \varphi)$, can be obtained from the relation $\psi_-(\bar\varphi) = \psi^*_+(\varphi)$, so the eigenfunctions of the basis for the state of the boson fields in the symmetric universe turns out to be given by (\ref{PSI01}) with the replacements, $t \rightarrow -t$ and $\varphi_k \rightarrow \bar \varphi_k$. Therefore, if the scalar field $\varphi$ represents  the boson field $W^-$ in one of the universes, then, $\bar \varphi$ represents the boson field, $\bar W^- = W^+$, in the symmetric universe. The decay of the Higgs into the boson $W^+$ and $W^-$ can then be produced separately in the two symmetric universes.

Then, one can make the hypothesis that the intermediate gauge boson $W^+$ and $W^-$ are created in different universes, or at least at different rates in the two universes, without violating the global matter-antimatter asymmetry, an appealing scenario that is also suggested in \cite{Faizal2014, Boyle2018}. It is not mandatory that the asymmetry is complete but a small asymmetry in the decay of the Higgs into the $W^+$ and $W^-$ bosons in the two universes would eventually derive into an asymmetry in the production of fermions in the two universes due to the different decays of the $W^\pm$ bosons into fermions (see, (\ref{D1}-\ref{D3})). In the universe in which the boson $W^+$ predominates there would be an excess of the up quark with respect to the up antiquark, and accordingly, there would be an excess of protons over antiprotons, and matter would therefore dominate over antimatter. From the global picture of the two correlated universes the total amount of matter is always balanced with the total amount of antimatter so there is no global matter-antimatter asymmetry. One would expect a whole range of matter-antimatter distributions in the pairs of universes of the whole multiverse. Most of them are probably filled with an equal amount of matter and antimatter in each universe. After recomposing, those universes will eventually evolve into radiation dominated universes with no matter content. However, some of them will be created with the matter-antimatter asymmetry needed to form galaxies and planets, and eventually life in a universe like ours.

\begin{acknowledgments}
I thank Prof. M. Dabrowski for suggesting the anthropic principle for the matter-antimatter asymmetry in the whole multiverse, and to the members of the Szczecin Cosmology Group for their kind hospitality at the University of Szczecin, where part of this work was conceived. This work was supported by Comunidad de Madrid (Spain) under the Multiannual Agreement with UC3M in the line of Excellence of University Professors (EPUC3M23), in the context of the 5th. Regional Programme of Research and Technological Innovation (PRICIT). 
\end{acknowledgments}




\bibliography{../bibliography}

\end{document}